\documentclass[prl,preprint,showpacs]{revtex4}
\usepackage{graphicx}
\usepackage{pstricks}
\usepackage{amssymb}

\begin{document}

\title{Lanczos-based Low-Rank Correction Method for Solving the Dyson Equation in Inhomogenous Dynamical Mean-Field Theory}


\author{\firstname{Pierre} \surname{Carrier}}
\author{\firstname{Jok M.} \surname{Tang}}
\author{\firstname{Yousef} \surname{Saad}}
\affiliation{Minnesota Supercomputing Institute and Department of Computer Science \& Engineering, University of Minnesota, Minneapolis, MN 55455}
\email{carrier@cs.umn.edu}

\author{\firstname{James K.} \surname{Freericks}}
\affiliation{Department of Physics, Georgetown University, Washington, D.C., 20057}

\date{\today}

\begin{abstract}
Inhomogeneous dynamical mean-field theory  has been employed to solve many interesting strongly interacting problems from transport in multilayered devices to the properties of ultracold atoms in a trap. The main computational step, especially for large systems, is the problem of calculating the inverse of a large sparse matrix to solve Dyson's equation and determine the local Green's function at each lattice site from the corresponding local self-energy.
We present a new efficient algorithm, the Lanczos-based low-rank algorithm, for the calculation of the inverse of a large
sparse matrix which yields this local (imaginary time) Green's function.
The Lanczos-based low-rank algorithm is based on  a domain decomposition
viewpoint, but avoids explicit calculation of  Schur complements and relies
instead on low-rank matrix approximations derived from the Lanczos
algorithm, for solving the Dyson equation.
We report at least a 25-fold improvement of performance compared to explicit decomposition (such as sparse $LU$)
of the matrix inverse.
We also report that scaling
relative to matrix sizes, of the low-rank correction method on the one hand
and domain decomposition methods on the other, are comparable.
\end{abstract}

\pacs{31.15.xr,  31.15.aq, 71.10.Fd}

\maketitle

\section{Introduction} \label{sec:Introduction}
The Dyson equation is at the heart of many important physical problems, taking its roots in many-body formalisms for the Green's function \cite{Kadanoff62, Economou06}.
The real-space version of the Dyson equation is of particular importance in the framework of inhomogeneous dynamical mean-field 
theory (IDMFT) \cite{Pott99, Freericks06, Tran06, Freericks09, Freericks10, Freericks10b}.
In this framework, the solution of the Dyson equation, for a given lattice (e.g., square or cubic ``optical'' lattice), is being matched 
self-consistently to the solution of an impurity problem. 
A fixed-point iteration ensures that the two local Green's functions (for the lattice and for the impurity problem) are equal to each other. 
The self-energy, $\Sigma_i$, defined at each lattice site $i$, and introduced into the Dyson equation, is used
for numerically matching the two problems.
Details on the IDMFT algorithm can be found in \cite{Tran06, Freericks06, Freericks09, Freericks10, Freericks10b, Carrier11}.

This paper focuses primarily on the first part of the IDMFT numerical method, the Dyson equation solver, 
which constitutes the most computationally intensive part of the IDMFT algorithm.
Details on the second part, the impurity solvers, applied to the Fermi-Bose Falicov-Kimball model \cite{Falicov69, Brandt89, Ziegler06, Iskin09},
can be found in Ref.~\cite{Freericks09, Freericks10b, Carrier11}, as it is used for our numerical simulations shown below.

The matrix form of the Dyson equation (along the imaginary time axis) used in any IDMFT implementation, is given by \cite{Economou06, Freericks06}:
\begin{equation}
\textbf{G}_{ij}( i \omega_k) = \left\{ [\textbf{G}_{ij}(i\omega_k;U=0)]^{-1} 
                                - \Sigma_i( i \omega_k)\mbox{\boldmath{$\delta$}}_{ij} \right\}^{-1} \equiv \left\{ \textbf{H}_{ij}( i \omega_k) \right\}^{-1}.
\label{Dysonmatrix}
\end{equation}
The matrix $[\textbf{G}_{ij}( i \omega_k;U=0)]$ corresponds to 
the noninteracting Green's function, whose inverse includes a  kinetic- energy term 
representing the nearest-neighbor hopping 
from site $j$ to site $i$ on a square (or cubic) lattice, plus the Matsubara frequency along the imaginary time 
axis, $ i \omega_k$, on the diagonal $\textbf{H}_{ii}$, and the local potential which involves the sum of the chemical potential minus the local trapping potential; that is, we have  $[\textbf{G}_{ij}( i \omega_k;U=0)]^{-1}=t_{ij}+(i\omega_k+\mu-V_i)\delta_{ij}$.  It is noninteracting in the sense that 
it excludes the many-body  potential energy terms, represented instead by the self-energy, $\Sigma_i( i \omega_k)$.
Note that the self-energy varies from site to site, and is therefore \emph{inhomogeneous} \cite{Tran06}.
The explicit values of $\Sigma_i( i \omega_k)$ on each lattice site $i$ are deduced from an impurity solver, 
not described here \cite{Brandt89, Zlatic01, Iskin09, Carrier11} for each lattice site.
For convenience, we have defined $\textbf{H}$ in eq.~(\ref{Dysonmatrix}) as, $\textbf{G} \equiv \left\{ \textbf{H}\right\}^{-1}$,
to be used when describing the new algorithm below.
The matrix $\textbf{H}$ corresponds to a lattice model, and it is non-singular, of rank $N$.
It is a complex-symmetric matrix, with complex-diagonal elements 
and real-valued off-diagonal elements, called hopping terms.
A similar expression for the Dyson equation can also be deduced along the real frequency axis, $i\omega_k\rightarrow \omega+i\delta$ \cite{Freericks06}. Note that the off-diagonal elements in the matrix $\textbf{ H}$ are extremely sparse because the hopping is restricted to nearest neighbors, so there are typically only $Z$ nonzero terms, where $Z$ is the coordination number of the lattice (4 for a square lattice and 6 for a simple cubic lattice).
In IDMFT, the quantity of interest is essentially the diagonal of the inverse, $\textbf{G}_{ii}$,
which is then matched to the impurity Green's function \cite{Carrier11} inside a self-consistency loop.
Therefore, the main task of the numerical method developed below is to determine 
$$
diag(\textbf{G}) \equiv diag\left(\left( \textbf{H}\right)^{-1}\right)
$$
from eq.~(\ref{Dysonmatrix}).
\section{The Lanczos-based Low-rank Correction Algorithm} \label{sec:Lanczos}
The new algorithm that solves the diagonal of the inverse of the matrix 
in eq.~(\ref{Dysonmatrix}) essentially combines two general approaches of solvers,
one iterative and one direct.
Those two approaches are: the Lanczos scheme \cite{Simon} and the domain decomposition \cite{Smith96}.
We first briefly recall the Lanczos algorithm and its important elements needed for the rest, then show an important example of 
domain decomposition, and describe finally the new
Lanczos-based low-rank correction algorithm based on this example of domain decomposition.

\subsection{Standard Lanczos}
The standard Lanczos algorithm is a special case of the Arnoldi algorithm applied to real-symmetric or Hermitian matrices \cite{SaadBook}.
Let assume first that $\textbf{H}$ is real-symmetric.
The fundamental 3-terms recurrence of the Lanczos algorithm is given by \cite{Parlett}
$$
\beta_{j+1}\textbf{q}_{j+1} = \textbf{H} \textbf{q}_{j} - \alpha_j \textbf{q}_{j} - \beta_j \textbf{q}_{j-1}, \hspace{5pt} \mbox{for} j=1,\dots, N.
$$
This recurrence requires only one ``matrix-vector'' operation per Lanczos iteration, corresponding to the following operation: 
\begin{equation}
\textbf{u}_j \leftarrow \textbf{H} \textbf{q}_{j},
\label{eq:matvec}
\end{equation}
where $\textbf{u}_j$ is a temporary variable holding the resulting matrix-vector product.
In matrix form, the Lanczos recurrence gives rise to a tridiagonal matrix $T_j \equiv tridiag(\beta_j, \alpha_j, \beta_{j+1})$, at the $j^{th}$ Lanczos iteration.
Once $j=N$, the rank of the matrix $\textbf{H}$, we obtain the exact relation
\begin{equation}
\textbf{H} = \textbf{Q}_N \textbf{T}_N \textbf{Q}_N^{T}.
\label{QNTNQN}
\end{equation}
It would be possible in practice to determine $\textbf{H}^{-1}$ from this above matrix decomposition (See for example Ref.~\cite{Sidje}).
However, for very large $N$ this procedure is prohibitive, due to the necessity to use and store \emph{all} $N$
Lanczos vectors, in either a partial or 
full re-orthogonalization \cite{Parlett} implementation.

As just stated, this Lanczos scheme applies only to real-symmetric (or Hermitian) matrices \cite{Parlett}.
When $\textbf{H}$ is complex-symmetric, as in IDMFT, the standard Lanczos scheme needs to be modified.
Freund \textit{et al.}\ \cite{Freund92} have shown that applying the Lanczos algorithm to complex-symmetric matrices can be done by
replacing \emph{everywhere} in the algorithm (including re-orthogonalization) the usual inner product by an indefinite inner product defined as
a transposition only, excluding the conjugation, i.e.: $(\textbf{q},\textbf{u})_{\mathbb{C}} = \textbf{q}^T \textbf{u}$,
where $\textbf{q}$ and $\textbf{u}$ are any two complex vectors.
The corresponding indefinite ``norm'' (inside quotation marks, because it does not satisfy fundamental properties of a norm)
is: $||q||_{\mathbb{C}} = \sqrt{(\textbf{q},\textbf{q})_{\mathbb{C}}}$, the principal square root.
Note that the resulting indefinite ``norm'' can be a complex number, and thus result in a possible
breakdown, or division by zero, in the algorithm (when $||q||_{\mathbb{C}}$ may become null for two non-null vectors).
This potential breakdown is taken care of by allowing for a few extra iterations, making sure that the null space can be reached within a sufficient number of
Lanczos steps, as further described below in the context of the Lanczos-based low-rank algorithm.
\begin{figure}[tb]
   \centering
   \includegraphics[width=3.0in]{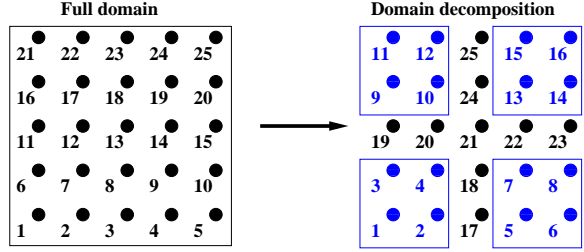}
   \caption{A 2-dimensional example of a domain, $\Omega$, discretized with $n = 5^2$ equidistant grid points. 
            The left part shows a conventional ordering of grid points; the
            right side shows a domain decomposition ordering, where grid points in the four subdomains are numbered first, 
            followed by the interface points.
           }
   \label{fig:ddm-ord}
\end{figure}

\subsection{Standard Domain Decomposition}
Domain decomposition is a very broad method \cite{Smith96}.
We describe here only one application, used next in the Lanczos-based low-rank algorithm.
Consider now two different site labelings of a square lattice depicted in Fig.~\ref{fig:ddm-ord}. 
A straightforward labeling of the nearest-neighbor hopping matrix, as part of the local Green's function 
$\textbf{G}_{ij}(U=0)$ for lattice sites $j$ to $i$ of eq.~(\ref{Dysonmatrix}),
is shown on the left of Fig.~\ref{fig:ddm-ord}, where lattice sites are
labeled consecutively from bottom-left to top-right. 
This labeling corresponds to a unique domain.
For hopping occuring only between nearest-neighbors, this straightforward numbering leads to a
penta-diagonal matrix (similarly to a 5-point Laplacian operator), in two-dimensions.
Of course, a multitude of different labelings can be constructed. 
Consider another important example, shown on the right of Fig.~\ref{fig:ddm-ord}, where
a decomposition into four subdomains of the nearest-neighbor lattice sites has been devised.
The corresponding matrix $\textbf{H}$, containing four subdomains, is no longer penta-diagonal, but has now the following form:
$$
\textbf{H} =
\left(\begin{array}{cccc|c}
\textbf{B}_1 & 0 & 0 & 0 & \textbf{E}_1 \\
0 & \textbf{B}_2 & 0 & 0 & \textbf{E}_2 \\
0 & 0 & \textbf{B}_3 & 0 & \textbf{E}_3 \\
0 & 0 & 0 & \textbf{B}_4 & \textbf{E}_4 \\ \hline
\textbf{E}^T_1 & \textbf{E}^T_2 & \textbf{E}^T_3 & \textbf{E}^T_4 & \textbf{C}
\end{array} \right).
$$
The four blocks, $\textbf{B}_i$, $i=1,\dots, 4$, can then be grouped into a block diagonal matrix $\textbf{B}$,
and similarly for the $\textbf{E}$ (and $\textbf{E}^{T}$):
\begin{equation}
\textbf{H} =
\left(\begin{array}{cc}
\textbf{B} & \textbf{E}\\ 
\textbf{E}^T & \textbf{C}
\end{array} \right).
\label{eq:domdec}
\end{equation}
The Schur complement
\begin{equation}
\textbf{S} \equiv \textbf{C} - \textbf{E}^T\textbf{B}^{-1}\textbf{E}
\label{eq:Schur}
\end{equation}
is then defined such that the matrix in eq.~(\ref{eq:domdec}) can be re-written as a product of
lower and upper block-triangular matrices \cite{SaadBook}, as:
$$
\textbf{H} =    \left( \begin{array}{cc} \textbf{I}       & \textbf{0} \\
                                \textbf{E}^T\textbf{B}^{-1} & \textbf{I} \end{array} \right)
       \left( \begin{array}{cc}  \textbf{B}      & \textbf{E} \\
                                 \textbf{0}      & \textbf{S} \end{array} \right),
$$
where $\textbf{I}$ is the identity matrix of proper dimension. 
The rank $m$ of the Schur complement matrix \textbf{S} is equal to the number of interface points.
In the example of Fig.~\ref{fig:ddm-ord} on the right, the interface points labeled from 17 to 25 give a rank $m=9$ for $\textbf{S}$,
for an entire matrix $\textbf{H}$ of rank $N=25$.

The main expression giving the inverse of $\textbf{H}$ based on the domain decomposition of Fig.~\ref{fig:ddm-ord},
can now be deduced from the above product of matrices. The resulting
inverse takes the form:
\begin{eqnarray}
\textbf{H}^{-1} & = & \left( \begin{array}{cc} \textbf{B}^{-1} & -(\textbf{B}^{-1}\textbf{E})\textbf{S}^{-1} \\
                                      \textbf{0}      & \textbf{S}^{-1} \end{array} \right)
              \left( \begin{array}{cc} \textbf{I} & \textbf{0} \\
                                       -(\textbf{E}^T\textbf{B}^{-1}) & \textbf{I} \end{array} \right) \nonumber \\
       & = & \left( \begin{array}{lc} \textbf{B}^{-1} & \textbf{0} \\
                                      \textbf{0}      & \textbf{0} \end{array} \right) + 
             \left( \begin{array}{cc} (\textbf{B}^{-1} \textbf{E})\textbf{S}^{-1} (\textbf{E}^T\textbf{B}^{-1}) & -(\textbf{B}^{-1} \textbf{E})\textbf{S}^{-1} \\
                                      -\textbf{S}^{-1} (\textbf{E}^T\textbf{B}^{-1})  & \textbf{S}^{-1}  \end{array} \right) \nonumber \\
        & = & \left( \begin{array}{lc} \textbf{B}^{-1} & \textbf{0} \\
                                  \textbf{0}      & \textbf{0} \end{array} \right) 
                                  + \left( \begin{array}{c} (\textbf{B}^{-1} \textbf{E}) \\ -\textbf{I} \end{array} \right) 
              \textbf{S}^{-1} \left( (\textbf{E}^T\textbf{B}^{-1}) \hspace{10pt} -\textbf{I} \right).\label{eq:domain_method}
\end{eqnarray}
Standard domain decomposition methods for finding $\textbf{H}^{-1}$ (or only its diagonal) are based on solving eq.~(\ref{eq:domain_method}), 
with the inverse of the Schur complement obtained from
the triple matrix product in eq.~(\ref{eq:Schur}).
This expression involves several matrix multiplications plus an inversion of the full Schur complement matrix $\textbf{S}$.
Efficient methods which drop non-significant terms (when only the diagonal is needed) 
are essential for obtaining a rapid algorithm \cite{Cauley07, Lin10, JokDD},
at the price of higher complexity.
Recalling that the dimension of the matrix $\textbf{S}$ equals that of $\textbf{C}$, i.e., 
the number of interface points $m$,
then using more domains leads to more interface points,
and consequently to a larger size Schur complement which is obviously more difficult to invert.
We have gathered at this point all the necessary ingredients for developing the Lanczos-based low-rank algorithm.

\subsection{Combination of Lanczos with Domain Decomposition}
The Lanczos-based low-rank correction algorithm starts with an identical 
decomposition of the domain as  shown   on   the   right   of
Fig.~\ref{fig:ddm-ord},  leading  to eq.~(\ref{eq:domain_method})   for   the  inverse   $\textbf{H}^{-1}$.
It however avoids any explicit  calculation of  the inverse  of
the Schur complement.
The  algorithm  is obtained
from  eq.~(\ref{eq:domain_method})  by   first  grouping  the  inverse
$\textbf{H}^{-1}$  together  with  the blocks  $\textbf{B}^{-1}$ on the right-hand side of eq.~(\ref{eq:domain_method}),  and
consequently  defining $\textbf{X}$ as:
\begin{equation}
\textbf{X} \equiv \textbf{H}^{-1} - \left( \begin{array}{lc} \textbf{B}^{-1} & \textbf{0} \\
                                  \textbf{0}      & \textbf{0} \end{array} \right)
=
\left( \begin{array}{c} (\textbf{B}^{-1} \textbf{E}) \\ -\textbf{I} \end{array} \right) \textbf{S}^{-1} \left( (\textbf{E}^T\textbf{B}^{-1}) \hspace{10pt} -\textbf{I} \right).
\label{eq:Schur1}
\end{equation}
The fundamental  idea of this new algorithm is based on  the fact that
the  right-hand  side of  eq.~(\ref{eq:Schur1})  has  rank $m$,  much
smaller  than $N$.  Consequently,  $\textbf{X}$ in eq.~(\ref{eq:Schur1}),
is  also of rank  $m \ll  N$.  To see this, recall that
the rank  of  $\textbf{S}$  corresponds  to  the  number  of
interface points  of the domain decomposition,
related to  $\textbf{C}$
as defined in eq.~(\ref{eq:domdec}). 
The matrix $\textbf{X}$ has, by construction, the same rank as $\textbf{S}$ (or $\textbf{C}$).
The diagonal of the  inverse of $\textbf{H}$ can thus  be determined
after $m \ll N$ Lanczos steps with the matrix $\textbf{X}$:
\begin{equation}
\textbf{X}= \textbf{Q}_m \textbf{T}_m \textbf{Q}_m^{T},
\label{eq:XQTQ0}
\end{equation}
i.e., instead of using the entire $N$ steps on the original matrix $\textbf{H}$, as in eq.~(\ref{QNTNQN}).
Beyond the m$^{th}$ Lanczos iteration the null space is reached
and $q_{m+1}\beta_{m} e_m^{*} \equiv 0$ --- so the expression in eq.
(\ref{eq:XQTQ0}) is \emph{exact}. 
In this situation,
we have the \emph{exact} following identity, after combining eq.~(\ref{eq:XQTQ0}) with eq. (\ref{eq:Schur1}):
\begin{equation}
\textbf{Q}_m \textbf{T}_m \textbf{Q}_m^{T} =  
\left( \begin{array}{c} (\textbf{B}^{-1} \textbf{E}) \\ -\textbf{I} \end{array} \right) \textbf{S}^{-1} 
\left( (\textbf{E}^T\textbf{B}^{-1}) \hspace{10pt} -\textbf{I} \right).
\label{Lanczos_equals_DD}
\end{equation}
Equation (\ref{Lanczos_equals_DD}) is the fundamental identity of the Lanczos-based low-rank correction algorithm.

In any standard Lanczos implementation \cite{Parlett}, the  matrix-vector  operation given in the expression (\ref{eq:matvec})
must now be replaced by a ``matrix-vector'' operation with $\textbf{X}$, as:
$$
\textbf{u}_j \leftarrow \textbf{X}\textbf{q}_j.
$$
This in fact translates as one system solve, one inversion of small block matrices, and an 
update based on the difference defined in eq.~(\ref{eq:Schur1}):
\begin{eqnarray}
\mbox{Solve}\hspace{5pt} \textbf{H} \textbf{v}_j & = & \textbf{q}_j, \label{Gv=q}\\
\mbox{Compute}\hspace{5pt} 
                         \textbf{w}_j & = & 
                         \left( \begin{array}{lc} \textbf{B}^{-1} & \textbf{0} \\
                         \textbf{0}      & \textbf{0} \end{array} \right) \textbf{q}_j, \label{Bw=q} \\
\mbox{Output:}\hspace{5pt}\textbf{u}_j &  \leftarrow & \textbf{v}_j - \textbf{w}_j. \label{eq:u=v-w}
\end{eqnarray}
Equations (\ref{Gv=q})--(\ref{eq:u=v-w}) replace the usual matrix-vector operation of standard Lanczos implementations \cite{Parlett}.
The initial problem of finding the inverse of a matrix has thus been replaced by one system solve per Lanczos step, for the
matrix $\textbf{H}$ in eq.~(\ref{Gv=q}).
In the example of Fig.~\ref{fig:ddm-ord} on the right, only 9 out of 25 possible Lanczos steps are required on $\textbf{X}$, and consequently, 
only 9 system solves based on eq.~(\ref{Gv=q}) are necessary. Computing the block diagonals in eq.~(\ref{Bw=q}) is relatively straightforward, 
since each block has relatively small size.

The diagonal of the inverse is finally obtained from the expression:
\begin{equation}
diag\left(\textbf{H}^{-1}\right) = diag\left(\begin{array}{lc} \textbf{B}^{-1} & \textbf{0} \\
                               \textbf{0}      & \textbf{0} \end{array} \right) + diag\left(\textbf{Q}_m \textbf{T}_m \textbf{Q}_m^{T} \right),
\label{diagGminus1}
\end{equation}
deduced from eq.~(\ref{eq:Schur1}) and eq.~(\ref{Lanczos_equals_DD}) for the diagonal terms only.
The diagonal of $\textbf{B}^{-1}$ is a constant matrix and needs to be calculated only once
at the very beginning of the Lanczos steps.
Standard domain decomposition given by eq.~(\ref{eq:domain_method}) and using eq.~(\ref{eq:Schur}) 
requires two matrix-matrix multiplications for obtaining the inverse of the Schur complement, while in the 
Lanczos scheme, the $\textbf{T}_m$ matrix is simply a tridiagonal matrix $tridiag(\beta_m, \alpha_m, \beta_{m+1})$, which
implies  that the  product  of the  three matrices  $diag[\textbf{Q}_m
\textbf{T}_m \textbf{Q}_m^{T}]$ in eq.~(\ref{diagGminus1}) is much simpler, and corresponds to the following
algebraic expression obtained by updating only three Lanczos vectors:
$$
diag\left(\textbf{Q}_m \textbf{T}_m \textbf{Q}_m^{T} \right) = \beta_i (\textbf{q}_{i-1} \odot \textbf{q}_{i})  
                                 + \alpha_i (\textbf{q}_i \odot \textbf{q}_i) 
                                 + \beta_{i+1} (\textbf{q}_{i} \odot \textbf{q}_{i+1}).
$$
The symbol $\odot$ stands for term-by-term vector product.

Note that, as mentioned before in the Lanczos Section, 
the algorithm of Freund \textit{et al.} \cite{Freund92} applied to complex-symmetric matrices potentially give
null eigenvalues in the matrix $\textbf{T}_m$, once the
$m$ Lanczos steps are done.
This fact translates as the null space being visited \emph{before} the termination of the $m$ Lanczos iterations.
It implies that once $m$ Lanczos iterations are performed, 
the Lanczos vector basis does not yet form a complete basis of the low-rank matrix $\textbf{X}$, and
therefore eq.~(\ref{eq:XQTQ0}) still contains an error term that is not yet \emph{exactly} null.
We use $\alpha_j$, $j=1, \dots, m$, the diagonal terms of $\textbf{T}_m$, as a stopping criteria;
as soon as $\alpha_j$ becomes null then the null space is reached.
In practice, a few extra iterations beyond $m$ are allowed.
In the IDMFT implementation, we find that no more than 5 or 10 extra iterations are required, for $m$ of the order of 100 for instance.
Other stopping criteria, such as $||\textbf{q}||< \epsilon$, using the ${\cal L}_2$ norm on the Lanczos vectors, or
by eventually computing all eigenvalues of $\textbf{T}_m$ when $m$ is relatively small,
can be used for determining when the null space is reached beyond the $m$ Lanczos steps, and then stop.

Note finally that the Lanczos-based low-rank algorithm is well-suited to recursive calls, as for any domain decomposition implementation.
Various levels of recursion of the domain in the Lanczos-based low-rank correction method are done by alternating the number of domains,
$(d_x, d_y)$, respectively along the $x$ and $y$ directions,
using for instance the series $(d_x, d_y) =(2,1), (1,2), (2,1), \dots$
We find that this  approach is the optimal recursion, reducing both the number of interface
points and the number of necessary recursions.

\section{Numerical Simulations} \label{sec:Numerical}
We evaluate next the performance of the new algorithm, applied to IDMFT for a two-dimensional Fermi-Bose Falicov-Kimball model Hamiltonian
\cite{Ziegler06, Iskin09, Freericks09, Freericks10}.
In IDMFT, multiple inversions of eq.~(\ref{Dysonmatrix}) need to be performed, i.e., for each Matsubara frequency included in the calculation, $ i \omega_k$ for $k=0$ up to $k_{max}$, and 
at each self-consistency iteration of the IDMFT loop (where the impurity and the Dyson solutions 
are matched, as briefly described in the Introduction). 
More details on the parallel IDMFT algorithm and its performance can be found in Refs.~\cite{Freericks09, Freericks10, Carrier11}.
Note that a ``probing'' method for inverting eq.~(\ref{Dysonmatrix}), that takes advantage of 
diagonal dominancy for some matrices with large Matsubara frequencies $ i \omega_k$, has been implemented and is also 
discussed in Refs.~\cite{JokProbing, Carrier11}.
However, as the temperature parameter $T$ of the simulation diminishes, i.e., the ultimate
goal in ``ultracold'' simulations \cite{Freericks09,Freericks10, Freericks10b}, the matrices tend to become indefinite.
The matrices $\textbf{H}$ are also mostly indefinite in the real-time Green's function problem.
Therefore, we focus in this short paper on the computer performance of the Lanczos-based low-rank algorithm for the inversion of general matrices,
including indefinite matrices.

The original version of the IDMFT code \cite{Freericks09} used LAPACK routines,
where a decomposition into $\textbf{H}=\textbf{LU}$ was  then
completed with a full inversion, $\textbf{G} \equiv \textbf{H}^{-1} = \textbf{U}^{-1}\textbf{L}^{-1}$,
for finally extracting the diagonal.
These complex-valued LAPACK routines, however, make use of \emph{dense} matrix techniques.
For the sake of comparison, 
we compare instead the performance of the Lanczos-based low-rank implementation 
to that of a
preconditioned-GMRES \cite{SaadBook} solver that finds the solutions of the following systems of equations,
using sparse techniques:
\begin{equation}
\textbf{H} x^i =  {e}^i,\hspace{7pt} \mbox{for} \hspace{5pt} i=1, N,
\label{eq:n3method}
\end{equation}
where the ${e}^i$'s are Euclidian vectors. 
Each vector  $x^i$, solution of  the above system, corresponds  to the
columns  of $\textbf{G}=\textbf{H}^{-1}$;  
the diagonal  of the  inverse  is then
extracted  from  the  $x_i^i$ components.  
We note that the performance of this rather expensive algorithm, is still better 
than using the $\textbf{U}^{-1}\textbf{L}^{-1}$ decomposition from LAPACK;
sparse solvers for eq.~(\ref{eq:n3method}) thus give a higher bound of performance to the dense $\textbf{U}^{-1}\textbf{L}^{-1}$ 
decompositions previously used \cite{Freericks09}.

The performances of the Lanczos-based low-rank method shown next were obtained from calculations on a Cray XT5 supercomputer (the machine ``Kraken'') 
located at the NICS, as part of the Teragrid.
The dimension of the matrices were set to $N= n^2 =101^2$, where $n$ represents the number of lattice sites in both the $x$ and $y$ directions. 
63 Matsubara frequencies, and their corresponding matrices, are calculated (on separate processors) for each of the self-consistency
steps. The accuracy of the final solution is one part in $10^{8}$ for all calculations.
A preconditioned-GMRES sparse solver, based on eq.~(\ref{eq:n3method}), takes at most 717.64 CPU seconds ($\sim$ 12 minutes) 
for finding \emph{one} of the 63 inverses of $\textbf{H}( i \omega_k)$, among all the possible
inversions occuring in the entire self-consistency iteration.
(It corresponds in fact to $\omega_0$, the smallest Matsubara frequency.)
This implies that if for example 20 self-consistency IDMFT iterations are necessary for achieving self-consistency, 
then the total time for the IDMFT loop is about 20 $\times$ 718 = 14,360 seconds, or $\sim$4 hours, when
a method based on eq.~(\ref{eq:n3method}) is used.
Using the Lanczos-based low-rank correction method, the exact same IDMFT calculation takes 
at most 28.66 CPU seconds for finding the same inverses.
Thus, the 20 iterations in the IDMFT loop discussed above
take a mere 20 $\times$  29 = 580 seconds, i.e., less than 10 minutes.
This corresponds to a dramatic increase in performance (irrespective of the type of matrices, indefinite or diagonally dominant),
i.e., at least a 25-fold increase in computational speed (the increase versus the dense LAPACK approach is much larger).

We report in addition that the scaling with the dimension $N$ of the matrices, between domain decomposition 
with an explicit calculation of the Schur complement based on eq.~(\ref{eq:domain_method}), and the Lanczos-based low-rank correction,
are equivalent \cite{Carrier11}.
They however have different prefactors (slightly higher in the Lanczos-based algorithm) due to different levels of optimization in the programs.
This implies that the performance of the Lanczos-based low-rank method is equivalent to that of calculating explicitly the Schur complement inverses, a method
much more complex in its implementation. 
This property will become important when considering in the near future 3-dimensional implementations of this algorithm.

\section{Conclusions and Perspectives} \label{sec:Conclusions}
The Lanczos-based low-rank correction method elegantly combines two important and usually distinct numerical methods 
for finding the diagonal of the inverse of complex symmetric sparse matrices:
the Lanczos iterative scheme, and the direct domain decomposition method.
We have shown how this new algorithm avoids the explicit computation of the Schur complement.
The algorithm transforms the inversion problem into a much smaller set of system solves, 
using a low-rank version of the original matrix to be inverted, and where the right-hand side vectors of the system solves are 
Lanczos vectors.
We used a 2-dimensional implementation of IDMFT applied to the Fermi-Bose Falicov-Kimball model Hamiltonian and showed that
the performance of this new algorithm corresponds to a 25-fold improvement, compared to other sparse methods.
The algorithm is currently being implemented inside a three-dimensional IDMFT Dyson solver applied to the Hubbard model \cite{Hubbard63}, and combined
with a continuous time quantum Monte-Carlo impurity solver to solve for system sizes $N$ on the order of a few million for direct comparison to experiment \cite{Esslinger10}.

\section{Aknowledgements}\label{sec:Acknowledgement}
We thank Daniel Osei-Kuffuor for his help with the preconditioners with the software ARMS.
We acknowledge Shuxia Zhang, David Porter, and the University of Minnesota Supercomputing Institute for support and resources.
This work is sponsored by NSF under grants OCI-0904587 and OCI-0904597, through TeraGrid resources,
under the American Recovery and Reinvestment Act (ARRA) of 2009.
J.K.F. also acknowledges the McDevitt endowment trust from Georgetown University.

\bibliography{LowRankLanczos_preprint}

\end{document}